\documentclass[final,5p,times,twocolumn]{elsarticle}
\usepackage{amssymb}
\usepackage{amsmath}
\usepackage{xcolor}
\usepackage{color,soul}
\usepackage{amsfonts}
\usepackage{algorithmic}
\usepackage{graphicx}
\usepackage{textcomp}
\usepackage{hyperref}
\usepackage{wrapfig,colortbl}
\usepackage{float}
\usepackage{multirow}
\usepackage{makecell}
\usepackage{booktabs}
\usepackage{enumitem}
\usepackage{bm}
\usepackage{caption}

\journal{Photoacoustics}

\begin{document}

\begin{frontmatter}

\title{Stochastic numerical head phantoms to enable virtual imaging studies of transcranial photoacoustic computed tomography}

\author[uiucece]{Hsuan-Kai Huang}
\author[uiucece]{Joseph Kuo}
\author[uiucbioe]{Seonyeong Park}
\author[utaustin,utbme]{Umberto Villa}
\author[caltech]{Lihong V. Wang}
\author[uiucbioe,uiucece]{Mark A. Anastasio}

\affiliation[uiucece]{organization={Department of Electrical and Computer Engineering, University of Illinois Urbana-Champaign},
            city={Urbana}, postcode={61801}, state={IL}, country={United States}
            }
\affiliation[caltech]{organization={Andrew and Peggy Cherng Department of Medical Engineering, Department of Electrical Engineering, California Institute of Technology},
            city={Pasadena}, postcode={91125}, state={CA},   country={United States}
            }
\affiliation[utaustin]{organization={Oden Institute for Computational Engineering and Sciences, The University of Texas at Austin},
            city={Austin}, postcode={78712}, state={TX}, country={United States}
            }
\affiliation[utbme]{organization={Department of Biomedical Engineering, The University of Texas at Austin},
            city={Austin}, postcode={78712}, state={TX}, country={United States}
            }
\affiliation[correspond]{organization={Department of Bioengineering, University of Illinois Urbana-Champaign},
            city={Urbana}, postcode={61801}, state={IL}, country={United States}
            }

\begin{abstract}
Transcranial photoacoustic computed tomography (PACT) is an emerging neuroimaging modality, but skull-induced aberrations can result in severe image artifacts if not compensated for during image reconstruction. The development of advanced image reconstruction methods for transcranial PACT is hindered by the lack of well-characterized, clinically relevant evaluation frameworks. Virtual imaging studies offer a solution, but require realistic numerical phantoms. To address this need, this study introduces a framework for generating ensembles of realistic 3D numerical head phantoms for virtual imaging studies. The framework uses adjunct CT data to create anatomical phantoms, which are then enhanced with stochastically synthesized vasculature and assigned realistic optical and acoustic-elastic properties. The utility of the framework is demonstrated through a case study on the impact of skull modeling errors on transcranial PACT image quality. By allowing researchers to assess and refine reconstruction methods meaningfully, the presented framework is expected to accelerate the development of transcranial PACT.
\end{abstract}


\end{frontmatter}

\section{Introduction}
\label{sec:intro}
Photoacoustic computed tomography (PACT) is a rapidly emerging imaging modality that exploits optical contrast and ultrasonic detection to achieve imaging with sub-millimeter resolution and centimeter-scale depth~\cite{wang_photoacoustic_2012, wang_prospects_2008, wang_practical_2016}. It can provide functional information based on rich optical contrast, making it highly promising for neuroimaging applications. In small animals such as mice, PACT has been applied to visualize brain structures and monitor hemoglobin dynamics for functional brain studies~\cite{wang_noninvasive_2003, li_real-time_2010, nasiriavanaki_high-resolution_2014}. Translation to human imaging has recently been demonstrated in patients who underwent hemicraniectomy, a procedure involving partial skull removal, producing brain vascular and functional images comparable to those from magnetic resonance imaging~\cite{na_massively_2022}. These findings underscore the potential of PACT for human neuroimaging.

However, accurate image reconstruction in transcranial PACT of subjects with intact skulls remains challenging.
The skull possesses significantly higher compressional wave speeds than soft tissue and, as elastic solids, also supports shear wave propagation. As the photoacoustic pressure field induced within the brain propagates through the skull, the compressional wave partially converts to shear waves at the first interface, and the shear wave component partially converts back to compressional waves at the second~\cite{schoonover_compensation_2011}. These mode conversions, combined with the heterogeneous acoustic and elastic properties of the skull, introduce strong aberrations in the pressure field. Such aberrations cause severe artifacts, degrading image quality if not adequately compensated for during image reconstruction~\cite{schoonover_compensation_2011}. 

Conventional model-based image reconstruction methods can effectively compensate for skull-induced aberrations~\cite{huang_aberration_2012, poudel_iterative_2020}, but they require accurate knowledge of the skull's acoustic and elastic properties. 
One promising research direction is the development of reconstruction methods that can mitigate skull-induced aberrations without accurate prior knowledge of the skull properties~\cite{poudel_joint_2020, huang_fast_2025, huang_gradient-free_2025}. Such methods must also be computationally tractable and accurate for routine use in studies involving human subjects.

A significant challenge hindering the development of advanced, effective image reconstruction methods for transcranial PACT is the lack of well-characterized, clinically relevant evaluation frameworks. 
Because \textit{in-vivo} imaging data lack ground truth, they are generally unsuitable for quantitative evaluation. Physical phantoms, while providing controlled imaging targets, are often relatively simplistic and do
not accurately describe the spatially variant physical and physiological parameters of the human head.
It is also generally difficult or costly to produce a large number of physical phantoms that convey realistic variability in the object properties.
This is problematic because using only a small number of phantoms can bias image quality assessment and may not accurately reflect the average image quality corresponding to a cohort of subjects.
In addition, data-driven approaches, such as Bayesian-based~\cite{goh_coupled_2018} and learning-based techniques~\cite{huang_fast_2025}, typically require large datasets for training. 

Virtual imaging studies, also known as computer simulation studies, offer a promising alternative to physical phantom experiments when realistic numerical phantoms and accurate computational models of the data acquisition process are employed~\cite{abadi_virtual_2020, park_stochastic_2023, li_3-d_2022, angla_transcranial_2023}. In virtual imaging studies of transcranial PACT, numerical head phantoms (NHPs) are required. The NHPs should incorporate accurate biomechanical models of the skull to simulate skull-induced acoustic aberrations. These NHPs should also include vasculature and other soft tissue structures relevant to transcranial PACT.
However, existing head phantoms developed for use in simulating other imaging modalities~\cite{segars_4d_2010, iacono_mida_2015, wheelock_high-density_2019, fang_mesh-based_2010} often lack these features.
Even when the skull is included, it is frequently oversimplified as a single, uniform layer, neglecting the heterogeneity and porosity of individual plates and layers~\cite{motherway_mechanical_2009}. 
Relying on these simplified models can lead to inaccurate simulations of pressure wave propagation~\cite{jones_comparison_2015, kyriakou_full-wave_2015, jiang_numerical_2020, robertson_effects_2018}, fundamentally limiting the fidelity of transcranial PACT studies.

This work aims to develop a framework for stochastically generating ensembles of realistic three-dimensional (3D) NHPs from adjunct imaging data to facilitate virtual imaging studies of transcranial PACT. In this framework, an anatomical head phantom is generated based on adjunct computed tomography (CT) data, which consists of a skull model that can be customized to reflect different degrees of anatomical heterogeneity and the corresponding acoustic and elastic properties. The anatomical phantom also contains stochastically synthesized vasculature in both the scalp and cortical regions.
Realistic optical and acoustic-elastic properties are then stochastically assigned to each tissue type, yielding the corresponding optical and acoustic head phantoms.
To demonstrate the usefulness of the proposed framework, a case study is conducted to investigate the impact of skull modeling errors on image quality in transcranial PACT. The generated NHP are virtually imaged, and the resulting pressure data are used for image reconstruction under different skull modeling assumptions.

The remaining sections of this article are organized as follows. Section \ref{sec:bg} provides a review of wave propagation in acoustic-elastic media and existing NHPs that are potentially relevant to transcranial PACT. The proposed framework for producing ensembles of realistic NHPs for use in virtual imaging studies of transcranial PACT is detailed in Section \ref{sec:method}. Illustrative examples of NHPs generated using this framework are presented in Section \ref{sec:example}. Section \ref{sec:study} presents the case study that explores the impact of skull modeling mismatches on image reconstruction quality. Finally, Sections \ref{sec:discussion} and \ref{sec:conclusion} present a discussion of the results and the conclusion of this work.

\section{Background}
\label{sec:bg}
\subsection{Transcranial PACT imaging physics}
\label{sec:bg_physics}
The skull is an elastic solid that can significantly attenuate and aberrate recorded PACT pressure data~\cite{huang_aberration_2012, na_photoacoustic_2021}. Approximating the skull as a fluid medium simplifies the wave propagation model but can lead to substantial artifacts~\cite{schoonover_compensation_2011,white_longitudinal_2006,mitsuhashi_forward-adjoint_2017}.
Wave propagation can be more accurately modeled by use of the elastic wave equation. This approach accounts for both longitudinal and shear waves within a heterogeneous medium consisting of soft tissue and skull~\cite{mitsuhashi_forward-adjoint_2017, poudel_iterative_2020, luo_full-wave_2024}.

The salient aspects of 3D wave propagation in an acoustic-elastic medium are reviewed here. 
While some approaches rely on domain decomposition~\cite{luo_full-wave_2024}, applying separate wave equations to elastic and acoustic mediums, the approach reviewed here applies a single set of equations uniformly across the entire domain.
Specifically, in a heterogeneous isotropic acoustic-elastic medium, the propagation of the stress tensor $\boldsymbol{\sigma}^{ij}(\bm{r},t)$ and acoustic particle velocity $\mathbf{v}(\bm{r},t)$ at location $\bm{r}\in \mathbb{R}^{3}$ and time $t\in [0, \infty)$ can be modeled by the following elastic wave equations~\cite{alterman_propagation_1968, virieux_p-sv_1986, bolt_seismology_2012, madariaga_modeling_1998}:
\begin{flalign}
\label{eq:elastic}
\begin{aligned}
    \partial_t \mathbf{v}(\bm{r},t) &+ \alpha (\bm{r}) \mathbf{v}(\bm{r},t) = \frac{1}{\rho(\bm{r})}(\nabla \cdot \boldsymbol{\sigma} (\bm{r},t))\\
    \partial_t \boldsymbol{\sigma} (\bm{r},t) &= \lambda (\bm{r}) \mathbf{tr} (\nabla \mathbf{v} (\bm{r},t))\mathbf{I}\\
    &+ \mu (\bm{r}) \left( \nabla \mathbf{v} (\bm{r},t) + \nabla \mathbf{v} (\bm{r},t)^T \right),
\end{aligned}
\end{flalign}
subject to the initial conditions,
\begin{equation}
\label{eq:init}
    \boldsymbol{\sigma}_0(\bm{r}) \equiv \boldsymbol{\sigma}(\bm{r},t)|_{t=0}=-p_0(\bm{r})\mathbf{I},\ \mathbf{v}(\bm{r},t)|_{t=0}=0.
\end{equation}
Here, $\rho(\bm{r})$ denotes the medium's density distribution, and $\lambda(\bm{r})$ and $\mu(\bm{r})$ represent the distributions of Lam\'e's first and second parameters. Finally, $p_0(\bm{r})$ represents the photoacoustically-induced initial pressure distribution, which is assumed to be non-zero only in the fluid medium. The longitudinal and shear wave speeds are given by $c_l(\bm{r}) \equiv \sqrt{(\lambda(\bm{r})+2\mu(\bm{r}))/\rho(\bm{r})}$ and $c_s(\bm{r}) \equiv \sqrt{\mu(\bm{r})/\rho(\bm{r})}$, respectively. For a fluid medium, $\mu(r)=0$ and therefore $c_s(\bm{r})=0$. The operator $\mathbf{tr} (\cdot)$ in \eqref{eq:elastic} calculates the trace of a matrix, and $\mathbf{I} \in \mathbb{R}^{3\times3}$ is the identity matrix. The acoustic attenuation is modeled as a diffusive absorption term, $\alpha (\bm{r}) \mathbf{v}(\bm{r},t)$, which is proportional to $\mathbf{v}(\bm{r},t)$ with the ratio being the frequency-independent acoustic diffusive absorption coefficient $\alpha(\bm{r})~$\cite{pinton_attenuation_2012}. This can be valid when the transducer frequency band is narrow. Frequency-dependent attenuation should be considered~\cite{poudel_three-dimensional_2019} when using a broadband transducer. The explicit derivation of these equations is detailed in the literature~\cite{poudel_three-dimensional_2019, pratt_medical_2019}.

\subsection{Previously developed numerical head phantoms}
\label{sec:bg_nhps}
Numerical head phantoms must satisfy several requirements for use in realistic virtual imaging studies of transcranial PACT. First, they should capture the multilayered anatomical structure of the head, including scalp, skull, and cortical tissues, with explicit representation of cortical vasculature. Beyond geometry, these phantoms must be endowed with realistic optical~\cite{wang_biomedical_2012} and viscoelastic properties (mentioned in Section \ref{sec:bg_physics}) to ensure accurate simulation of both light transport and acoustic-elastic wave propagation. Additionally, they should incorporate stochastic variation in features such as skull viscoelastic properties and vascular distribution to reflect inter-subject diversity. Finally, to ensure translational relevance, stochastic phantoms should be anchored to adjunct imaging data where possible. However, existing NHPs fail to fulfill these requirements.

Several frameworks for producing 3D anatomical NHPs have been proposed~\cite{segars_4d_2010, iacono_mida_2015}, primarily to facilitate virtual imaging studies of modalities such as CT, magnetic resonance imaging (MRI), and diffuse optical tomography. By assigning the required optical and acoustic properties, these phantoms could potentially be used in transcranial PACT studies. However, they often lack the specific anatomical details critical for this application. For instance, cortical vascular anatomy, a primary imaging target in transcranial PACT, is often not well-represented~\cite{segars_4d_2010, iacono_mida_2015} or is even omitted in some frameworks~\cite{wheelock_high-density_2019, fang_mesh-based_2010}. When vasculature is derived from magnetic resonance angiography (MRA) data, it predominantly represents deeper brain vessels and largely neglects superficial cortical networks, which are the primary imaging targets in transcranial PACT. Although many of the reported works include detailed descriptions of brain regions (e.g., thalamus, hippocampus) and tissues (e.g., gray matter, white matter), this level of detail offers little benefit for current transcranial PACT applications due to restricted imaging depth and limited optical contrast of such structures.

Additionally, most of the phantoms feature an oversimplified skull model. Accurate skull modeling is an essential component of acoustic NHPs for computing pressure wave propagation through the skull~\cite{angla_transcranial_2023}. Studies have consistently shown that neglecting skull heterogeneity can compromise computation accuracy~\cite{jones_comparison_2015, kyriakou_full-wave_2015, jiang_numerical_2020, robertson_accurate_2017}. However, anatomical NHPs that are not specifically designed for modeling acoustic wave propagation often employ oversimplified, homogeneous skull models~\cite{segars_4d_2010, wheelock_high-density_2019, fang_mesh-based_2010}. 

By generating ensembles of phantoms that capture a wide range of anatomical variability, researchers can stress-test reconstruction methods and evaluate their robustness to real-world subject-dependent differences in skull viscoelastic properties and vascular networks. By systematically altering and evaluating tissue property assumptions, stochastic NHPs enable the identification of conditions that are most likely to induce artifacts, an investigation that cannot be performed on actual patients. 
Furthermore, using stochastic phantoms allows for large-scale \textit{in-silico} experiments that are infeasible in costly and ethically constrained human studies, thereby enabling the efficient exploration of numerous scenarios before \textit{in-vivo} validation.

In summary, while anatomically and physiologically realistic NHPs have been developed for several imaging modalities (e.g., CT, MRI, and positron emission tomography), there remains a clear need for a stochastic framework capable of producing ensembles of 3D NHPs based on a target skull subject and of comprehensively incorporating the essential anatomical, optical, and acoustic properties required for advancing transcranial PACT imaging research.

\section{Methods}
\label{sec:method}
To address the limitations described above, a new framework was developed to produce ensembles of 3D NHPs for virtual imaging studies of transcranial PACT. The NHPs in the generated ensemble contain skull geometry deduced from adjunct CT data for a given subject. Each realization of NHP incorporates variability in the cortical and scalp vasculature as well as in tissue-specific optical, acoustic, and elastic properties of the head. Each NHP consists of a tuple of \{anatomical NHP, optical NHP, acoustic-elastic NHP\}, collectively referred to as an NHP (without qualification). By use of the framework, ensembles of NHPs can be generated that share a common skull geometry and structure while varying in soft tissue geometry and the tissue-specific properties of the head. Repeating this procedure for different subjects with their own adjunct CT data yields ensembles that capture inter-subject variability in skull geometry and structure, which is relevant for cohort-based studies.

\begin{figure}[ht]
    \centering
    \includegraphics[width=0.9\linewidth]{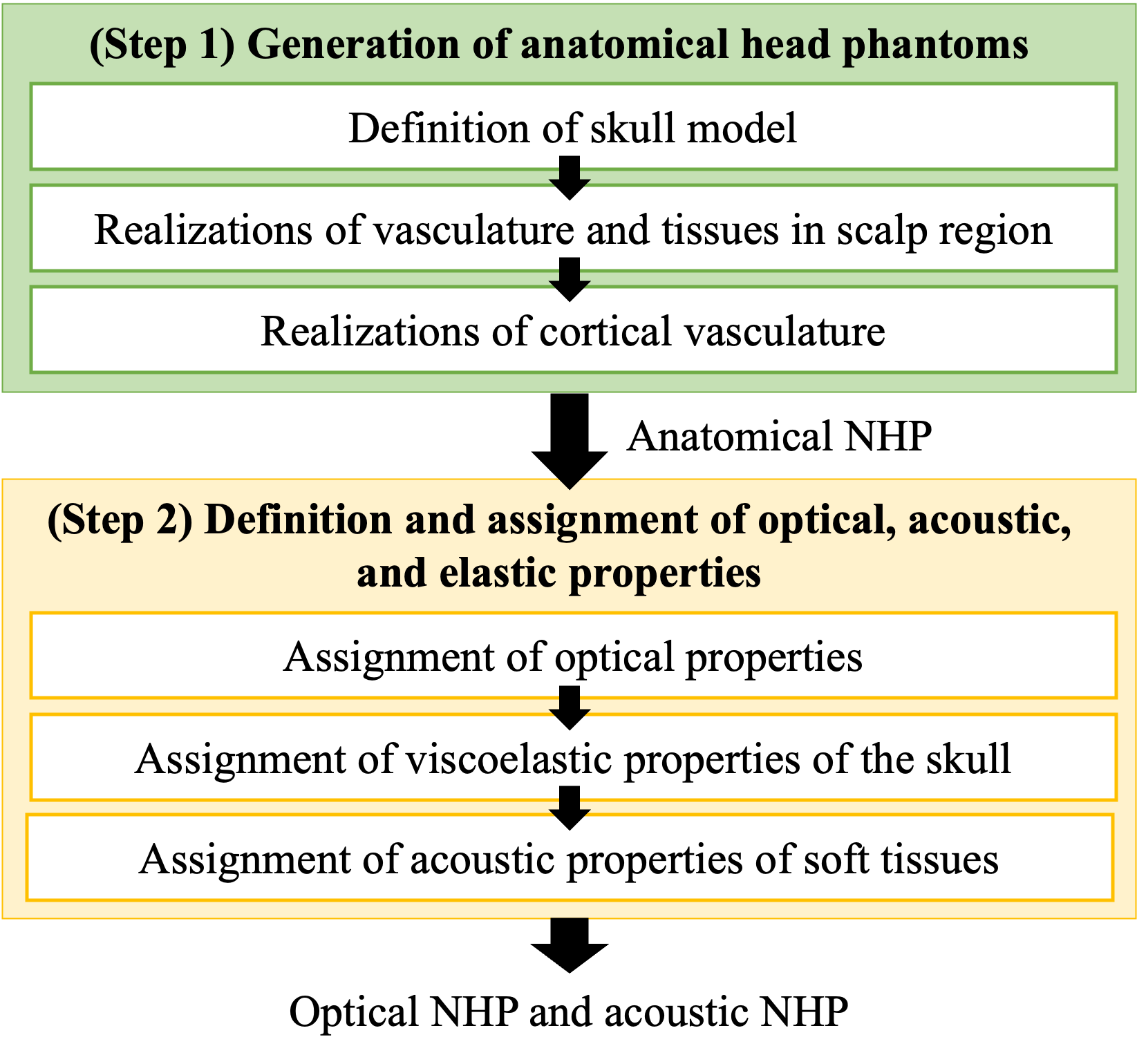}
    \caption{Workflow for generating NHPs for transcranial PACT. The two-step process begins with the modeling of the skull and the establishment of the anatomical NHP, and is finalized by the definition and assignment of optical, elastic, and acoustic properties.}
    \label{fig:workflow}
\end{figure}

The process of creating a realization of the stochastic NHP model consists of two main steps, as illustrated in Fig. \ref{fig:workflow}:

\noindent \textbf{Step 1}. An anatomical NHP is constructed to describe the geometry and structure of the head. The skull's geometry and composition are determined by segmenting adjunct CT data. The cortical and scalp vasculature are stochastically generated. This results in an ensemble of anatomical NHPs for a given subject, sharing 
a common skull but differing in soft tissue configurations.

\noindent \textbf{Step 2}. For each anatomical NHP generated in Step 1, optical, acoustic, and elastic properties are assigned to the skull and soft tissues. These properties can be assigned deterministically using nominal values from literature or stochastically using probabilistic models with nominal values as the distribution means. This results in a complete NHP set consisting of the anatomical NHP from Step 1 together with the corresponding optical and acoustic-elastic NHPs.

\subsection{Generation of anatomical NHPs}
The anatomical NHP describes the head anatomy and is defined by multiple tissue types. In this work, all tissue types are categorized into three distinct regions: skull, scalp, and cortical regions. The construction process begins with a skull model derived from adjunct imaging data, from which the geometry of the scalp and cortical regions is delineated. Within these regions, vasculature is stochastically synthesized. Details of skull geometry and vasculature synthesis are discussed below.

\subsubsection{Definition of skull geometry and components}
\label{sec:skull_geo}

\begin{figure}[ht]
    \centering
    \includegraphics[width=0.9\linewidth]{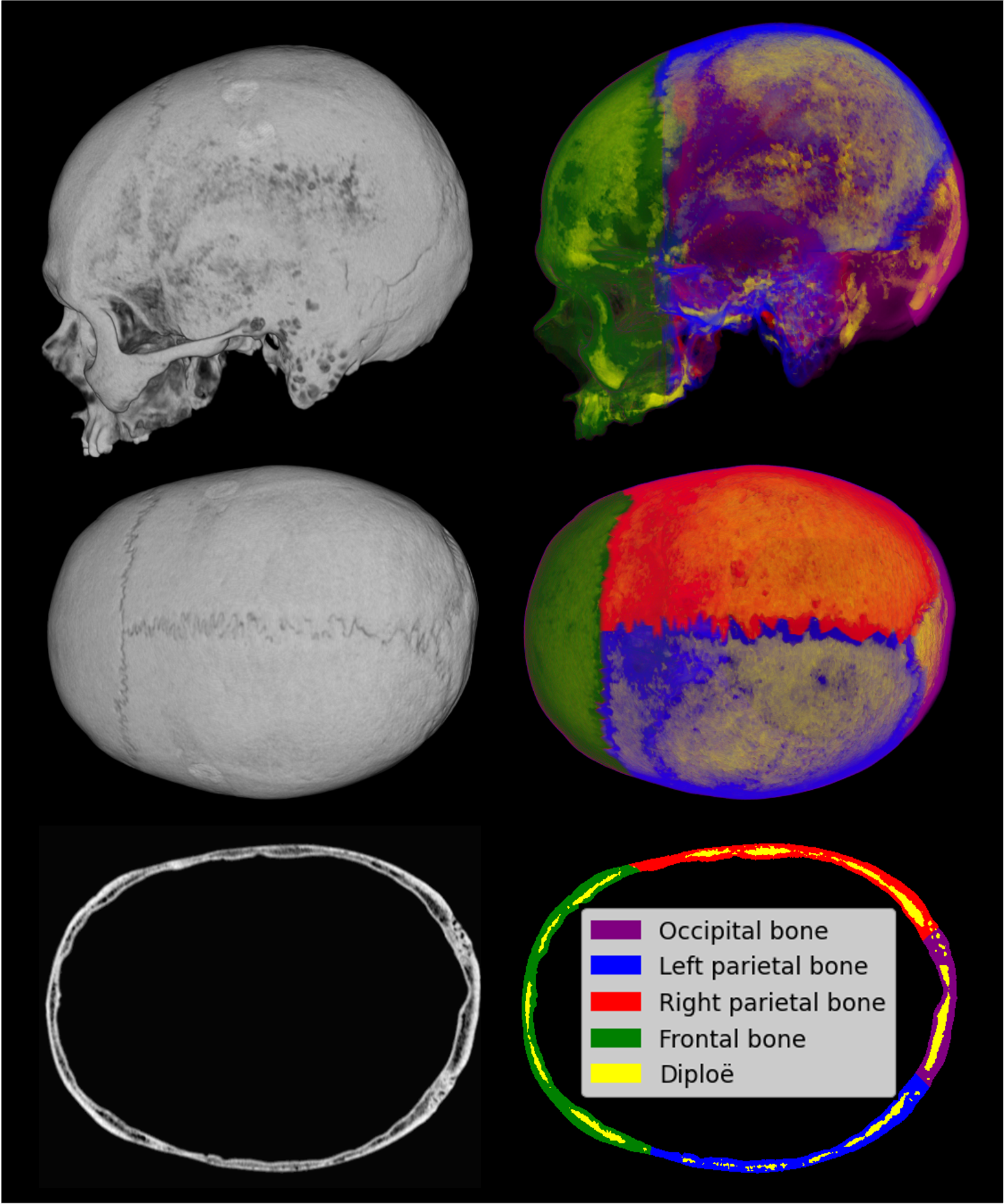}
    \caption{Segmentation results for establishing the skull model. The left panels show CT intensity projection images along the frontal (top) and longitudinal axis (center), and a cross-section in the transverse plane (bottom) of the skull. The right panels present the corresponding segmented skull plates and layers.}
    \label{fig:skull}
\end{figure}

To construct the skull structural model, adjunct CT images are segmented into multiple plates and layers. The model describes both the outer geometry of the skull (e.g., shape and thickness) and the spatial arrangement of its internal structures (e.g., plates and layers). This representation is essential for defining the anatomy of the NHP and for assigning the corresponding physical properties. For example, as shown in Fig. \ref{fig:skull}, a skull model was constructed from the adjunct CT data of a donated adult human skull~\cite{cao_single-shot_2023}. In this example, the skull was segmented into four plates (frontal, right parietal, left parietal, and occipital bones) and three layers (the diplo\"e layer, inner skull layer, and outer skull layer~\cite{marieb_essentials_2011}). These four plates are sufficient to describe the anatomy of the top of the head.
The three layers are defined to highlight the presence of the diplo\"e. For this work, the inner and outer skull plates are jointly considered during property assignment, as they are not necessarily anatomically separate.

The specific segmentation method used here is not a core part of the phantom generation framework but serves as a representative example. The soft tissue-skull interface was derived from the adjunct CT data using Yen's thresholding method~\cite{yen_new_1995}. The boundaries between skull plates were estimated by visual inspection of the CT images to infer the locations of the sutures. The boundaries between the skull plates and the diplo\"e were determined using Otsu's thresholding method~\cite{otsu_threshold_1979}.

This 4-plate 3-layer skull model enables the assignment of distinct parameters to each plate and layer. However, the properties within each component remain heterogeneous, particularly the pore structure within the diplo\"e layer. This level of heterogeneity is addressed in Section \ref{sec:ac_assign}.

\subsubsection{Realization of soft tissue structure in scalp regions}
The scalp was modeled as an approximately 4 mm-thick homogeneous tissue layer positioned above the skull~\cite{hori_thickness_1972}, omitting the thin layer of connective tissue between the skull and the scalp. Scalp vasculature was synthesized using a constrained constructive optimization (CCO) algorithm to mimic the angiogenesis process. The CCO implementation from the open-source Virtual Iterative Angiogenesis (VITA) library~\cite{talou_adaptive_2021} was used. 
The scalp vasculature were synthesized within the pericranium (1-3 mm) and aponeurosis (1-2 mm) layers~\cite{tajran_anatomy_2023}. Other vasculature-relevant parameters, including bifurcation angle, width, number of segments, cost function, and angle constraints, were set to the default values reported for the brain cortex vascularization case in the VITA library publication~\cite{talou_adaptive_2021}, which was considered the most relevant to scalp vessels among the examples provided.

\begin{figure*}[ht]
    \centering
    \includegraphics[width=0.9\linewidth]{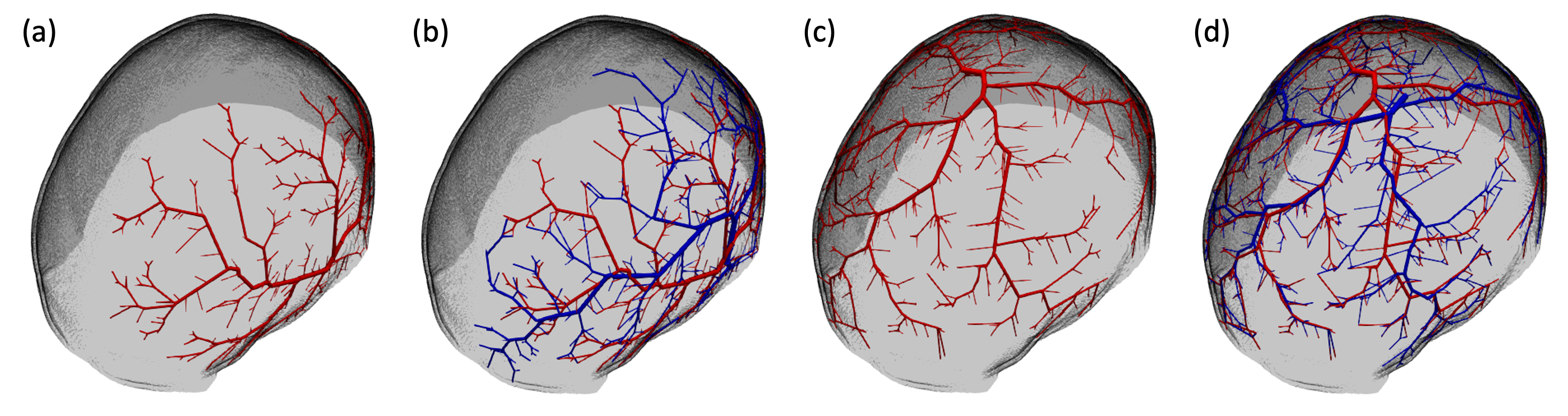}
    \caption{Four instances of scalp vasculature realization: (a) a vessel tree containing only the superficial temporal artery; (b) vasculature with both arteries and veins;     (c) a dense artery vessel tree originating from the top of the head; and (d) vasculature with both dense arteries and veins concentrated in a targeted region. The superficial temporal arteries and veins were synthesized based on realistic anatomy.}
    \label{fig:scalp_ves}
\end{figure*}

The framework allows for flexible adjustments of the scalp vasculature synthesis region. Four representative examples are shown in Fig. \ref{fig:scalp_ves}. In Fig. \ref{fig:scalp_ves} (a), the vasculature was synthesized to originate from the bottom left of the head to capture the realistic thickness variations of the superficial temporal artery.
In Fig. \ref{fig:scalp_ves} (b), both arteries and veins were generated, which enables assignment of oxygenation-dependent optical properties and facilitates multi-wavelength transcranial PACT for functional imaging applications. Figures \ref{fig:scalp_ves} (c) and (d) illustrate vessel synthesis originating from the top of the head to provide dense vasculature in that region. In Fig. \ref{fig:scalp_ves} (c), only arteries are shown, whereas Fig. \ref{fig:scalp_ves} (d) includes both arteries and veins. The resulting scalp vasculature was combined with the scalp to define the soft tissues in the scalp regions.

\begin{figure}[!b]
    \centering
    \includegraphics[width=0.9\linewidth]{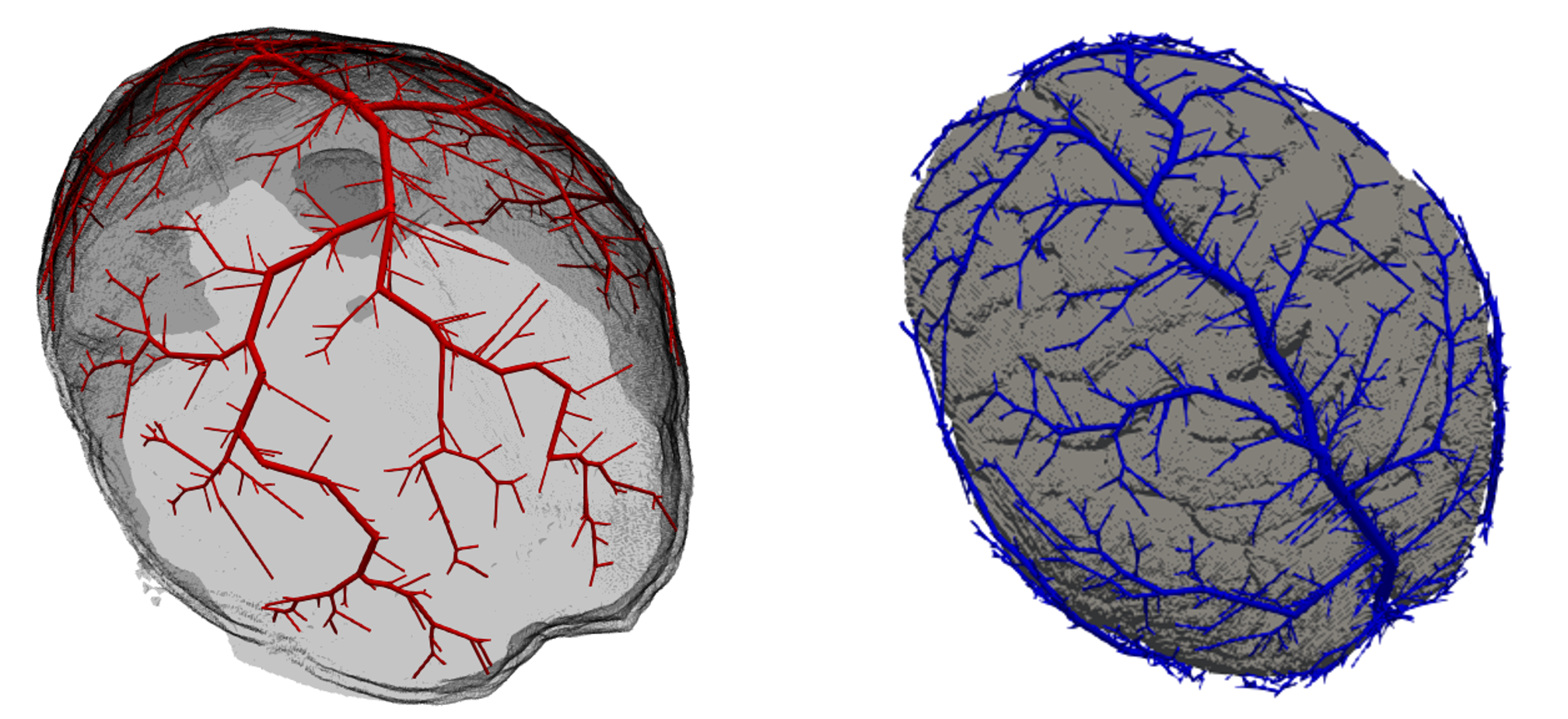}
    \caption{Examples of synthesized cortical vasculature. The left image shows dense cortical vasculature synthesized at the top of the head. The right image illustrates a model of the superior sagittal sinus, synthesized based on brain anatomy. This highlights the framework's capability to generate vasculature for diverse anatomical regions and imaging applications.}
    \label{fig:cor_ves}
\end{figure}

\begin{figure*}[ht]
    \centering
    \includegraphics[width=0.9\textwidth]{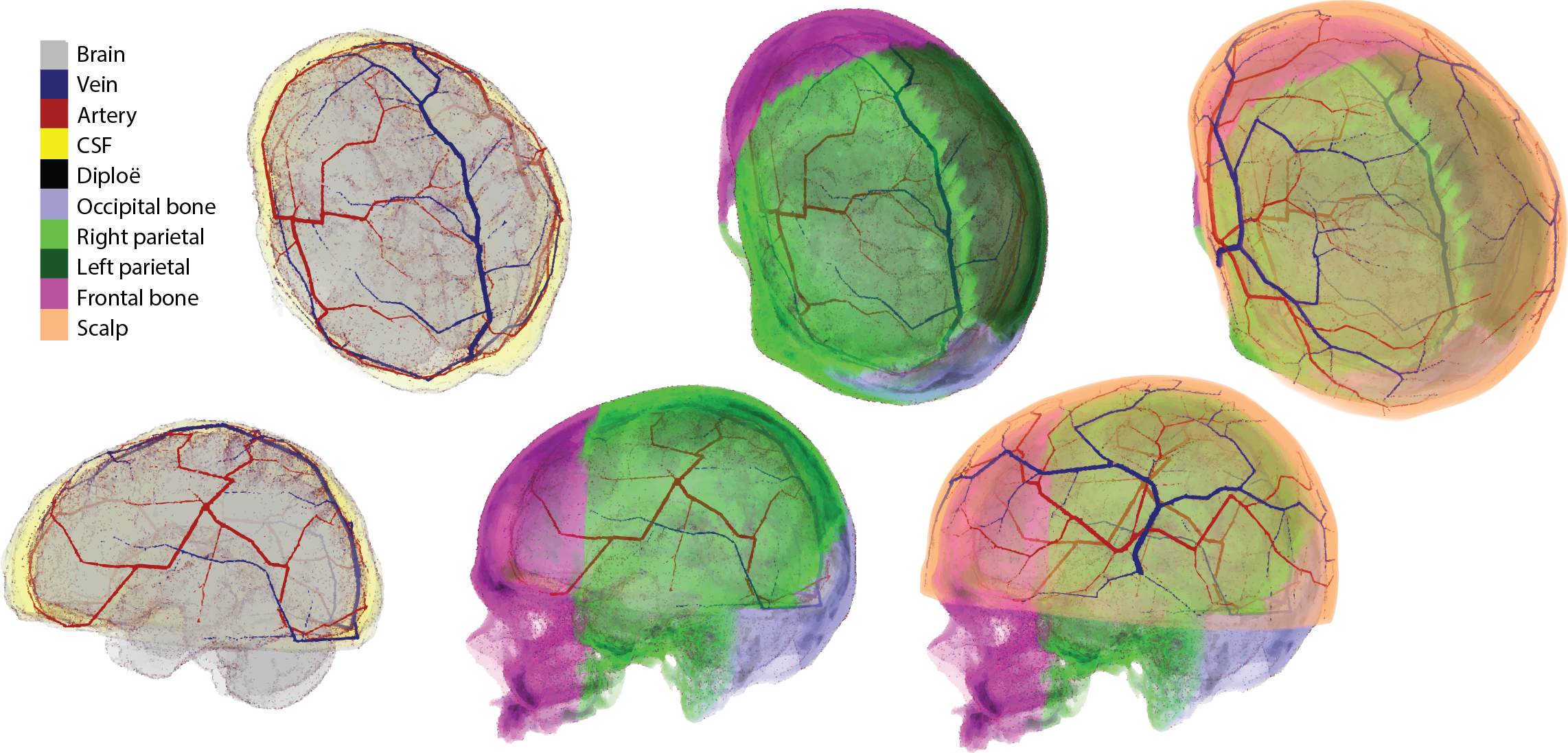}
    \caption{Progressive, component-by-component visualization of the head phantom, illustrating its layered anatomical structure. The left panels show the internal cortical vasculature and brain, while the middle and right panels sequentially add the skull and scalp, respectively, to form the complete external anatomy.}
    \label{fig:anatomical}
\end{figure*}

\subsubsection{Realization of soft tissue structures in cortical regions}
The cortical vasculature was synthesized using a workflow similar to that of the scalp vasculature. The region for synthesizing cortical vessels can be defined in two ways: either based on a fixed distance from the inner surface of the skull (within the arachnoid and dura mater layers) or by incorporating a brain atlas to reflect the underlying anatomy.

For the first approach, cortical vasculature can be synthesized within a region defined by a distance from the inner surface of the skull. This region corresponds to the arachnoid ($<$30 $\mu$m)~\cite{reina_structure_2010} and dura mater layers (0.4-1.4 mm thick)~\cite{zwirner_mechanical_2019}. The example on the left of Fig. \ref{fig:cor_ves} illustrates a dense vascular network synthesized from a vessel root located near the cranial vertex.

For the second approach, brain anatomy (e.g., the gyri and sulci) can be considered to synthesize more realistic vasculature, such as the middle cerebral artery and superior sagittal sinus. When \textit{in-vivo} CT data of the subject are available, this region can potentially be defined based on the brain anatomy inferred from the adjunct imaging data. In this work, CT data from an \textit{ex-vivo} human skull were used, and a brain atlas from an open-source library~\cite{wheelock_high-density_2019} was coregistered with the CT image of the donated skull using rigid transformation. As an example, the vasculature on the right of Fig. \ref{fig:cor_ves} illustrates a synthesized superior sagittal sinus, demonstrating the feasibility of incorporating brain anatomy into the synthesis process.

\subsubsection{Example of an anatomical NHP}
\label{sec:ana_example}
An example of a constructed anatomical NHP is illustrated in Fig. \ref{fig:anatomical}. This specific phantom includes scalp arteries and veins on both sides, the 4-plate 3-layer skull model, and cortical arteries and veins. For clarity, a hierarchical visualization of its components is provided. The NHP was designed to focus on the top region of the head, which is why soft tissue is absent from the lower half, and the temporal bones remain unsegmented. However, the NHP generation framework itself can be adapted to other imaging fields-of-view without significant modification.

\subsection{Assignment of optical, acoustic, and elastic properties}
\label{sec:prop_assign}
In Step 2 of the 3D NHP generation framework, anatomical NHPs are assigned optical, acoustic, and elastic properties to create optical and acoustic-elastic NHPs. To reflect the optical contrast that drives photoacoustic imaging, optical property assignments primarily focus on soft tissues and vasculature, while the skull is assumed to be optically homogeneous. Conversely, the assignment of acoustic and elastic properties focuses on the skull, as its complex acoustic-elastic characteristics must be accurately represented to realistically simulate transcranial wave propagation. Such realistic modeling is particularly important because it provides the foundation for developing and evaluating reconstruction methods that mitigate skull-induced aberrations and ensure high image quality in practical transcranial PACT.

\begin{table}[!b]
\small
\centering
\caption{Scattering coefficient parameters~\cite{jacques_optical_2013}.}
\label{tab:opt_scattering}
\begin{tabular}{l|c|c}
\toprule
Medium               & $a$\ [$\text{mm}^{-1}$] & $b$ \\
\midrule
Skull                & 0.95  & 0.141   \\ \hline
Gray matter          & 2.42  & 1.611   \\ \hline
Blood Vessels        & 2.20  & 0.660   \\ \hline
Scalp                & 4.78  & 2.453   \\
\bottomrule
\end{tabular}
\caption*{A reference wavelength is 500 nm.}
\end{table}

\begin{table}[h]
\small
\centering
\caption{Summary of optical properties of NHPs.}
\label{tab:opt_sum}
\begin{tabular}{l|c|c|c}
\toprule
Medium     & $\mu_{a}\,[\text{mm}^{-1}]$ 690/830/1064 nm  & $g$ & $n$ \\ 
\midrule
$\text{D}_2 \text{O}$    & 0.001/0.001/0.001~\cite{kedenburg_linear_2012} & 0.99   &  1.33~\cite{kedenburg_linear_2012} \\ \hline 
Skull      & 0.010/0.014~\cite{strangman_factors_2003}/0.022~\cite{bashkatov_optical_2006} & 0.944~\cite{bell_quantifying_2015}& 1.56~\cite{bell_quantifying_2015} \\ \hline 
CSF & 0.001/0.003~\cite{strangman_factors_2003}/0.013~\cite{kedenburg_linear_2012} & 0.99   &  1.33~\cite{kedenburg_linear_2012} \\ \hline 
Brain & 0.018/0.019~\cite{strangman_factors_2003}/0.055~\cite{yaroslavsky_optical_2002} & 0.89~\cite{jacques_optical_2013} & 1.40~\cite{biswas_vivo_2009} \\ \hline 
Artery     & 0.169/0.495/0.329 &  \multirow{2}{*}{0.976~\cite{meinke_empirical_2007}} & \multirow{2}{*}{1.35~\cite{sydoruk_refractive_2012}} \\ \cline{1-2} 
Vein       & 0.328/0.469/0.276 &  &  \\ \hline 
Scalp      & 0.049/0.043/0.021~\cite{bashkatov_optical_2005} & 0.87~\cite{ma_bulk_2005} & 1.36~\cite{ma_bulk_2005} \\ 
\bottomrule
\end{tabular}
\end{table}

\subsubsection{Assignment of optical properties}
\label{sec:opt_assign}
The optical NHPs were constructed from the anatomical NHP by assigning tissue-specific optical properties: absorption coefficient $\mu_a$, scattering coefficient $\mu_s$, anisotropy factor $g$, and refractive index $n$.
To facilitate functional imaging applications, the optical property values were provided at three optical wavelengths: 690 nm (deoxy-hemoglobin-dominant), 830 nm (intermediate wavelength), and 1064 nm (oxy-hemoglobin-dominant). The properties of gray matter were used to represent brain tissues excluding cortical vasculature, and the properties of water were used to represent cerebrospinal fluid (CSF). In addition, the properties of heavy water ($\text{D}_2 \text{O}$) were provided as a candidate acoustic coupling medium for transcranial PACT, which exhibits lower optical absorption than water, particularly at wavelengths of 830 nm and 1064 nm.

The scattering coefficient $\mu_s$ at each wavelength was computed from the reduced scattering coefficient $\mu'_s$ and anisotropy factor $g$, with the relation $\mu_{s} = \mu'_s/(1-g)$. The wavelength dependence of $\mu'_s$ was modeled with the power law~\cite{jacques_optical_2013}: $\mu'_s = a \big(\frac{\lambda}{500 \text{nm}}\big)^{-b}$, where parameters $a$ and $b$ are summarized in Table \ref{tab:opt_scattering}. Arterial and venous $\mu_a$ values were computed based on the extinction coefficients of oxy- and deoxy-hemoglobin~\cite{prahl_tabulated_1999, suzaki_noninvasive_2006}, total hemoglobin concentration in blood, and oxygen saturation~\cite{park_stochastic_2023}. The total hemoglobin concentration was randomly sampled from a uniform distribution $U(1860,2589)$, corresponding to the normal hematocrit range of 36-50\%~\cite{noauthor_blood_2022}. The oxygen saturation was randomly sampled from $U(95,99)$ for arteries and $U(75,84)$ for veins~\cite{nitzan_measurement_2008}. For other tissues, $\mu_a$ values were adopted from the literature~\cite{kedenburg_linear_2012, strangman_factors_2003, bashkatov_optical_2006, yaroslavsky_optical_2002, bashkatov_optical_2005}. Because $g$ and $n$ do not vary significantly over the near-infrared range (700 to 1100 nm)~\cite{jacques_optical_2013}, constant values from the literature~\cite{kedenburg_linear_2012, jacques_optical_2013, bell_quantifying_2015, biswas_vivo_2009, sydoruk_refractive_2012, ma_bulk_2005} were assigned to each tissue type regardless of the wavelength. The adopted optical property values, including the mean arterial and venous $\mu_a$ values, are summarized in Table \ref{tab:opt_sum}. These values were assigned to the corresponding tissue types in the anatomical phantoms to generate optical NHPs. The resulting optical NHPs can be used in optical fluence simulations to produce the photoacoustically induced initial pressure distributions, as detailed in Section \ref{sec:exvit}.

\begin{table}[ht]
\small
\centering
    \caption{Summary of the nominal viscoelastic property values.}
    \label{tab:elastic_properties}
    \begin{tabular}{l|c|c|c|c}
    \toprule
    \multirow{2}{*}{Medium} & $\rho$ & $c_{l}$ & $c_{s}$ & $\alpha$ \\
    & $[\text{kg}/\text{mm}^3]$ & $[\text{mm}/\mu \text{s}]$ & $[\text{mm} / \mu \text{s}]$ & $[1/\mu \text{s}]$ \\ 
    \midrule
    Water~\cite{white_longitudinal_2006}                & 1000.0 & 1.48  & 0     & 0      \\ 
    Entire skull~\cite{white_longitudinal_2006}         & 1800.0 & 2.80  & 1.40  & 0.60   \\ \hline
    Pure Bone            & 2520.0 & 4.00  & 2.00  & 0.90   \\ \hline 
    Frontal Plate        & 1944.4 & 3.04  & 1.52  & 0.54   \\ 
    Left Parietal        & 1920.5 & 3.00  & 1.50  & 0.55   \\ 
    Right Parietal       & 1928.2 & 3.01  & 1.51  & 0.55   \\ 
    Occipital Plate      & 2006.3 & 3.14  & 1.57  & 0.51   \\ 
    Diplo\"e             & 1441.3 & 2.21  & 1.10  & 0.75   \\ 
    \bottomrule
    \end{tabular}
\end{table}

\subsubsection{Assignment of viscoelastic properties of the skull}
\label{sec:skull_def}

As described in Section\ \ref{sec:bg_physics}, the viscoelastic properties considered here include density, bulk modulus, and shear modulus (for computing compressional and shear wave speeds), along with the absorption coefficient from the diffusive absorption model~\cite{moczo_finite-difference_2007} to describe acoustic absorption. Based on the skull segmentation in Section\ \ref{sec:skull_geo}, distinct properties can be assigned to each plate and layer. However, heterogeneity within each skull plate and layer requires careful consideration. Underestimating this heterogeneity can compromise the modeling accuracy of acoustic wave propagation and skull-induced aberrations~\cite{jones_comparison_2015, kyriakou_full-wave_2015, jiang_numerical_2020, robertson_accurate_2017}. On the other hand, precisely modeling subject-specific heterogeneity of skull properties remains significantly challenging, particularly in the diplo\"e layer. 

Several approaches have been reported to account for the heterogeneous skull properties, as summarized in Ref.~\cite{angla_transcranial_2023}. For example, the skull property heterogeneity can be estimated from an adjunct CT image by leveraging its relationship with viscoelastic properties~\cite{aubry_experimental_2003}. However, microstructural features in the diplo\"e layer (e.g., pore structure) generally cannot be fully captured because CT resolution is insufficient. Alternatively, homogenized skull properties can be assigned at low resolution, but this may introduce modeling errors~\cite{robertson_effects_2018, jones_comparison_2015, kyriakou_full-wave_2015, jiang_numerical_2020} 
unless the effects of the skull's microstructure are accurately modeled as equivalent macroscopic properties, which remains challenging and requires further exploration~\cite{liu_influence_2019, yousefian_frequency-dependent_2021}. 
The required level of heterogeneity in a model also depends on the acoustic frequency, as structures much smaller than the acoustic wavelength can often be treated as homogenized in modeling~\cite{angla_transcranial_2023}. 
In summary, the choice of skull viscoelastic property modeling should be aligned with the specific goals of a study, taking into account factors such as the resolution of adjunct CT data and acoustic wavelength.

\begin{figure}[ht]
    \centering
    \includegraphics[width=0.95\linewidth]{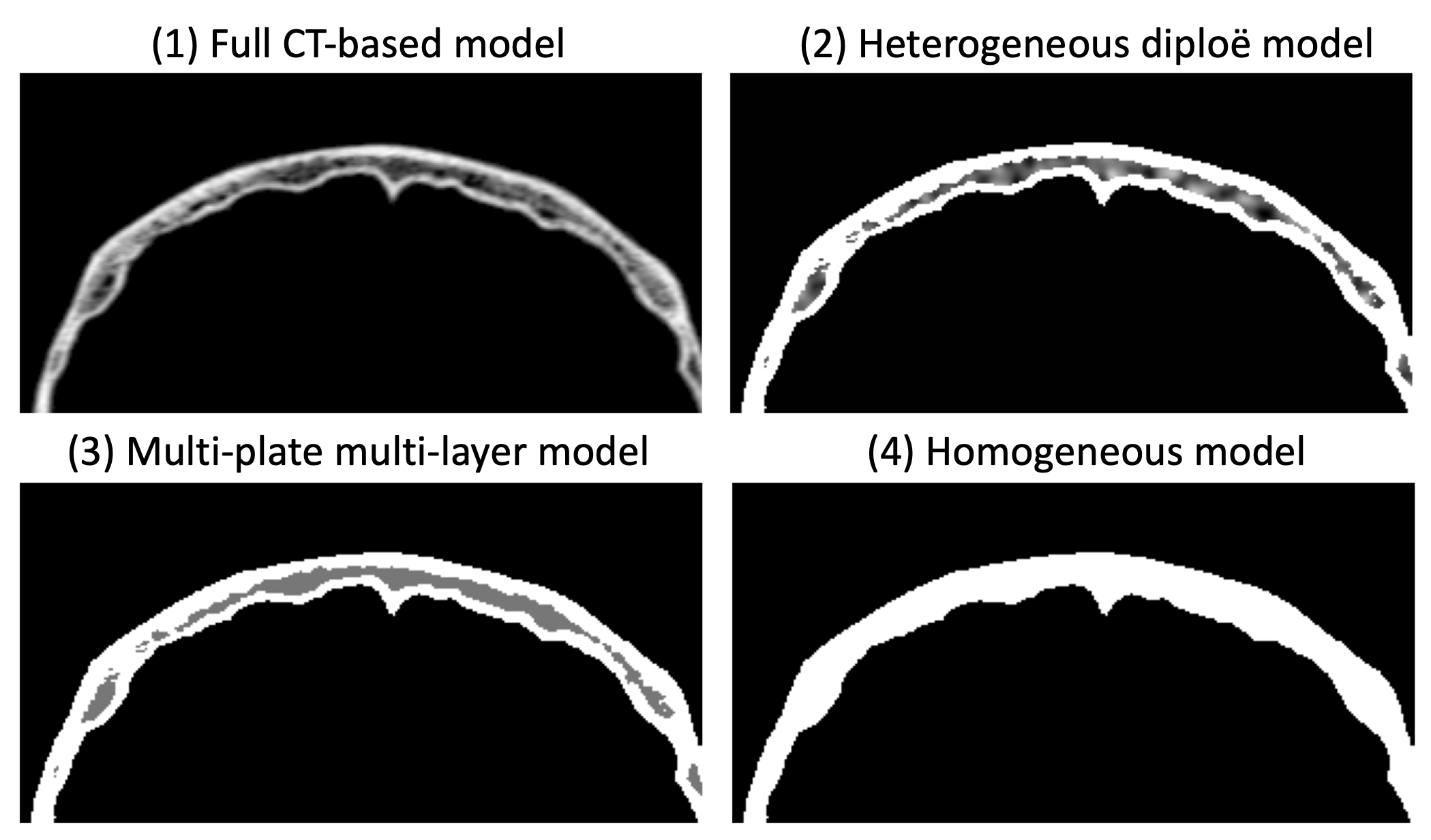}
    \caption{Cross-sections of the frontal bone area illustrating four distinct skull models with different levels of heterogeneity. These models can be selected to accommodate various research applications.}
    \label{fig:skulloption}
\end{figure}

To accommodate a variety of research scenarios, this framework provides four options for assigning the skull properties~\cite{angla_transcranial_2023}, each with a different level of heterogeneity: \textbf{Model 1}, a full CT-based skull model; \textbf{Model 2}, a heterogeneous diplo\"e skull model; \textbf{Model 3}, a multi-plate multi-layer skull model; and \textbf{Model 4}, a homogeneous skull model. Representative cross-sections of these models are shown in Fig. \ref{fig:skulloption}, illustrating the structural differences produced by each modeling approach. A comparison of these models is discussed in Section \ref{sec:discussion}. 

For simplicity, the property assignments described below are based on nominal values from the literature. However, the framework also allows the introduction of stochasticity 
to reflect inter-subject variability. For example, assigned values can be sampled from a normal distribution with the nominal value as the mean and a standard deviation of 5\% or 10\% of the mean. It should be noted that the true distribution of skull properties remains largely unexplored~\cite{auperrin_geometrical_2014}. The four skull models introduced above are described in more detail below.

\noindent \textbf{Model 1 - Full CT-based skull model:} In this approach, the spatial distributions of the viscoelastic properties are derived from a presumed linear relationship between viscoelastic properties and CT Hounsfield units (HU)~\cite{aubry_experimental_2003}. 
This model is applicable when high-resolution CT data are available and the property values of the subject-specific pure bone represented in the CT data (without pores) are known accurately. 
Specifically, the lowest HU values in the CT image are assumed to correspond to water, and the highest HU values to pure bone. Then, the CT image is scaled accordingly between the viscoelastic properties values of water and pure bone. The pure bone property values are adjusted to ensure that the average property value over the entire skull volume matches the nominal value reported in the literature~\cite{chang_development_2016, motherway_mechanical_2009, fry_acoustical_1978, auperrin_geometrical_2014, mcelhaney_mechanical_1970}. The values of water, pure bone, and the average skull volume (entire skull) properties used in this study are summarized in Table~\ref{tab:elastic_properties}.
Although this approach yields the most visually realistic skull models, the fundamental assumption of a linear relationship between CT and acoustic properties remains unverified~\cite{webb_measurements_2018}. 
Attempting to model fine-scale structural heterogeneity with an unverified parameter distribution may not provide a more realistic or predictive representation of skull-induced aberrations than a simplified, homogeneous model. 
This ambiguity in the property assignment limits the accuracy and applicability of the model. 
Further discussion on this topic is provided in Section \ref{sec:discussion}.

\noindent \textbf{Model 2 - Heterogeneous diplo\"e skull model:} This model incorporates voxel-level heterogeneous distributions of viscoelastic properties within the diplo\"e layer, while properties in individual skull plates (i.e., their inner and outer layers) remain piecewise constant. Two variants are considered for modeling the diplo\"e layer. \textbf{Model 2a} leverages the CT image in the same manner as Model 1 but only for the diplo\"e layer, whereas \textbf{Model 2b} introduces a stochastic representation of diplo\"e layer heterogeneity using a spatially correlated Gaussian random field~\cite{li_3-d_2022}. Model 2b can be used when high-resolution adjunct CT data are unavailable. It provides a means to reflect uncertainty in the heterogeneous properties of the diplo\"e layer while allowing porosity to be adjusted. In this study, the property values listed in Table~\ref{tab:elastic_properties} were assigned to each plate. These values were calculated by volume averaging applied to Model 1. For Model 2b, the Gaussian random field had a correlation length empirically set to 1.5 mm. The mean value of the field was matched to the corresponding HU of the used adjunct CT data, with a standard deviation of 1\% of this mean. The viscoelastic property distributions in the diplo\"e layer were then derived from this Gaussian random field, thereby emulating porosity. 

\begin{figure*}[ht]
    \includegraphics[width=0.95\textwidth]{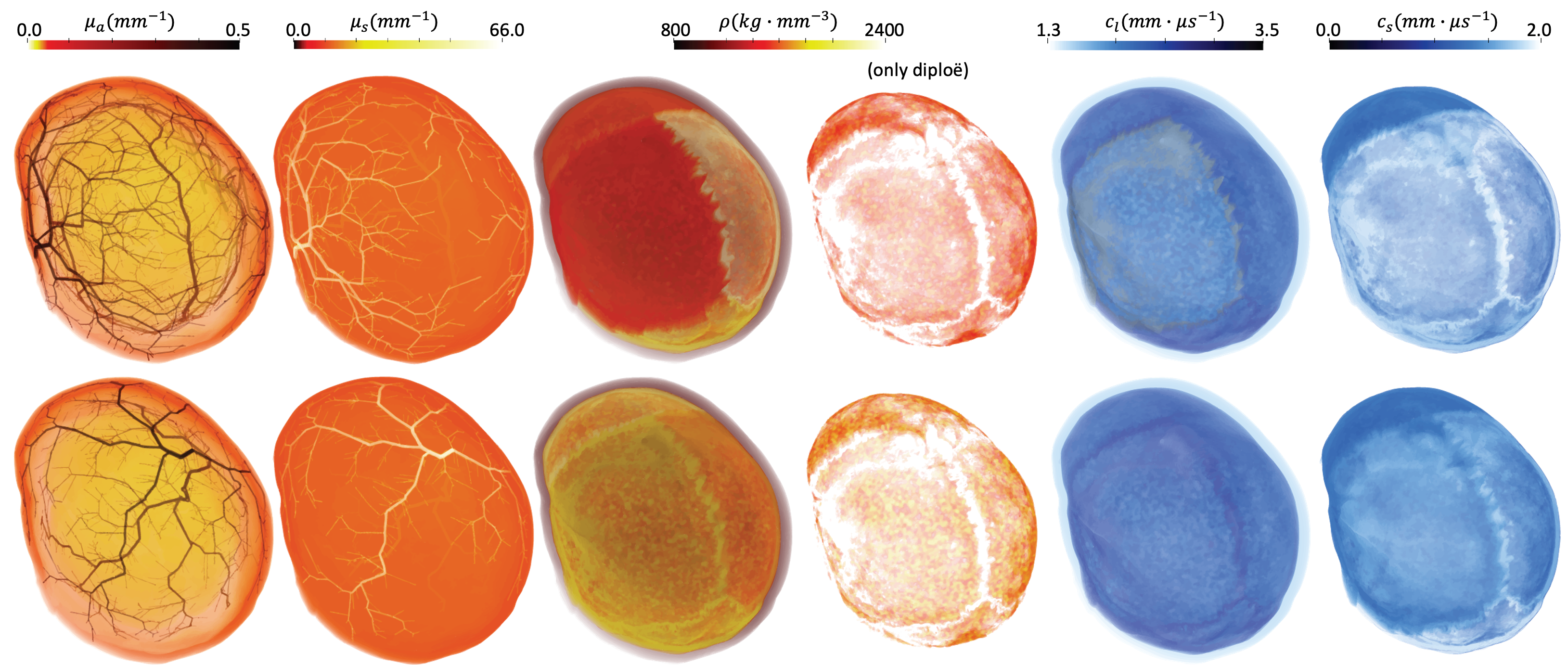}
    \caption{Two examples (top and bottom) of optical and acoustic-elastic NHPs. Illustrated properties, from left to right, include optical absorption coefficient $\mu_a$, optical scattering coefficient $\mu_s$, mass density $\rho$ of the entire head, mass density of the diplo\"e, compressional wave speed $c_l$, and shear wave speed $c_s$.}
    \label{fig:prop}
\end{figure*}

\noindent \textbf{Model 3 - Multi-plate multi-layer skull model:} In this model, the property assignment is piecewise constant for all individual plates and layers. The diplo\"e is represented as a homogeneous material with equivalent macroscopic properties~\cite{liu_influence_2019, yousefian_frequency-dependent_2021}, thereby neglecting the heterogeneous distributions arising from its porous microstructure. Thus, this model is suitable when such heterogeneity can be neglected, for example, at low acoustic frequencies~\cite{wintermark_t1-weighted_2014, fry_acoustical_1978}. The nominal values, derived by volume averaging the property distributions of Model 1, are summarized in Table~\ref{tab:elastic_properties} and can be used for this property assignment. 

\noindent \textbf{Model 4 - Homogeneous skull model:} This model treats the entire skull as a homogeneous elastic medium, thereby neglecting all internal heterogeneity. 
It is applicable when the skull's internal structure and viscoelastic properties cannot be determined and only its geometry can be estimated from low-resolution CT or other adjunct modalities, such as ultrashort echo time (UTE) MRI~\cite{deininger-czermak_evaluation_2022}. In this case, nominal values from the literature, summarized in Table \ref{tab:elastic_properties} (entire skull), can be used for property assignment. Because this model neglects the internal heterogeneity, it provides only a simplified approximation of wave propagation through the skull. 

\subsubsection{Assignment of acoustic properties of soft tissues}
\label{sec:ac_assign}
All soft tissues were assigned the acoustic properties of water at room temperature, as provided in Table \ref{tab:elastic_properties}. This assumption was made because the differences in property values between soft tissues are relatively small compared to those between soft tissue and the skull. In addition, the attenuation coefficient of the soft tissues was assumed to be zero. However, these assumptions can be readily relaxed. The shear wave speed of soft tissues was also set to zero, as they generally do not support the propagation of shear waves in the frequency range detected by ultrasonic transducers employed in PACT applications.

\section{Examples of NHPs and virtual imaging of transcranial PACT}
\label{sec:example}
To illustrate the characteristics of the NHPs generated by the framework, examples of optical and acoustic NHPs are presented in Section \ref{sec:exnhp}. Based on these examples, the photoacoustically-induced initial pressure distributions and the corresponding pressure measurements have been simulated and are described in Section \ref{sec:exvit}.  

\subsection{Examples of generated NHPs}
\label{sec:exnhp}
This section demonstrates the capabilities of the NHP generation framework by presenting examples of optical and acoustic phantoms. Figure \ref{fig:prop} shows representative property distributions, including the optical absorption and scattering coefficients, whole-head mass density, diplo\"e-layer mass density, and compressional and shear wave speeds (from left to right). The diplo\"e-layer mass density is separately illustrated to highlight the stochastically synthesized heterogeneity introduced to emulate porosity. The top row displays the optical and acoustic-elastic NHPs corresponding to the anatomical NHP in Fig. \ref{fig:anatomical}, in which cortical vasculature was synthesized using a brain atlas. The bottom row presents another set of phantoms, where cortical vasculature was synthesized at a fixed distance from the inner surface of the skull with its vessel root located near the cranial vertex. In both examples, the viscoelastic properties were assigned using Model 2b.

\begin{figure}[ht]
    \centering
    \includegraphics[width=0.45\textwidth]{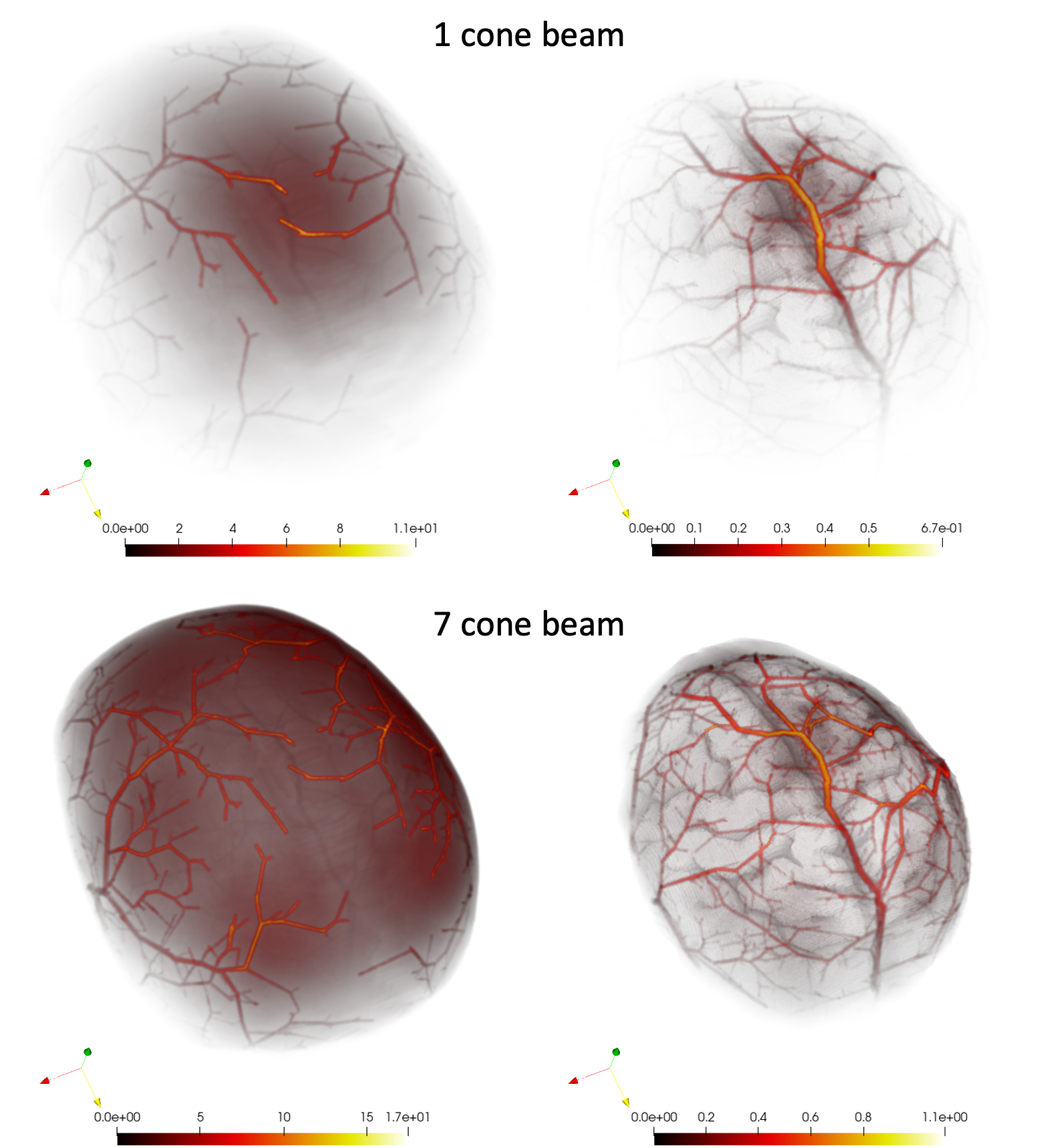}
    \caption{Simulated initial pressure $p_0$ distributions generated based on two distinct imaging systems (top and bottom). The left panels show the full $p_0$ distributions, while the right panels illustrate the same distributions with cortical field-of-view masking applied. Due to the exponential fluence decay with depth, the pressure amplitude of the cortical vasculature is nearly an order of magnitude smaller than that of the scalp vasculature.}
    \label{fig:p0phantom}
\end{figure}

\begin{figure*}[ht]
    \centering
    \includegraphics[width=0.95\textwidth]{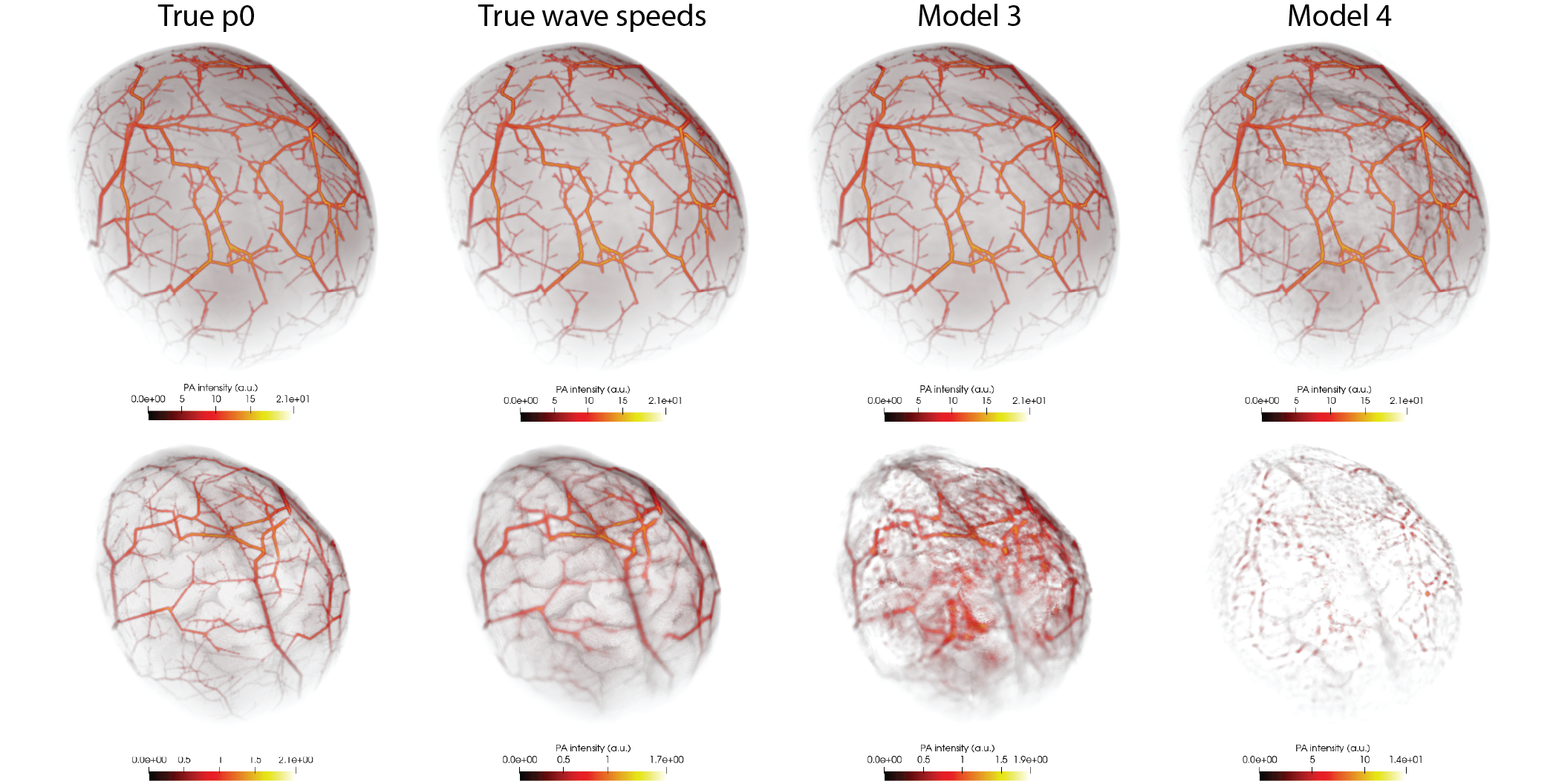}
    \caption{Visualization of true (first column) and estimated $p_0$ distributions (second to fourth columns) for the NHP in the case study. The top row displays the $p_0$ of the whole phantom, while the bottom row shows the $p_0$ within the cortical regions. When true wave speeds were used, the reconstructed cortical vasculature shows minor blurring due to the effects of noise and regularization. However, the $p_0$ reconstructed by assuming a homogeneous diplo\"e layer (Model 3) exhibits degradation from modeling errors, which becomes substantially more pronounced 
    when the entire skull was modeled as a homogeneous medium (Model 4).}
    \label{fig:casestudy}
\end{figure*}

\subsection{Examples of virtual imaging of transcranial PACT}
\label{sec:exvit}
To virtually image the generated NHPs, two virtual imaging systems were configured. The first system emulated a real-world imager~\cite{poudel_iterative_2020} with a downward-facing optical cone beam of 46$^\circ$ half-angle and 2.2 J pulse energy at a wavelength of 1064 nm. The second system augmented this beam with six identical side beams, evenly distributed in azimuth at a 45$^\circ$ polar angle, to ensure comprehensive illumination of the entire NHP. Both systems employed the same acoustic detection setup, modeled after a real-world 3D imager~\cite{aborahama_-aberration_2024}, consisting of a hemispherical measurement geometry with a 13 cm radius. The data acquisition period was 160 $\mu s$ with a 30 MHz sampling frequency, resulting in 4800 time points. 

Photon transport in biological tissues was simulated using the open-source MCX library~\cite{fang_monte_2009} to compute the induced initial pressure $p_0$ distributions. These $p_0$ distributions were then used as inputs to compute the PACT measurement data based on the elastic wave equation in Section~\ref{sec:bg_physics}. Acoustic wave propagation can be calculated based on several numerical schemes, such as the pseudo-spectral time-domain method~\cite{firouzi_firstorder_2011, treeby_modeling_2014}, the spectral element method~\cite{marty_towards_2024}, and the finite-difference time-domain (FDTD) method~\cite{mitsuhashi_forward-adjoint_2017}. In this work, the FDTD implementation based on the open-source Devito library~\cite{louboutin_devito_2019} was used to compute the measured pressure data. To improve the numerical stability in the wave propagation simulations, Gaussian spectral filtering was applied to the skull model with a correlation length of 1.5 mm. Electronic noise was modeled as additive, independent, and identically distributed Gaussian white noise, with the standard deviation (i.e., the noise level) set to 5$\%$ of the ensemble maximum amplitude of the simulated pressure data resulting from the cortical vasculature.

Virtual imaging was performed using the two system configurations described above with the NHP corresponding to anatomical NHP in Fig. \ref{fig:anatomical}. The resulting $p_0$ distributions are shown in Fig. \ref{fig:p0phantom}. Due to optical attenuation, the pressure amplitude of the cortical vasculature was nearly an order of magnitude lower than that of the scalp vasculature. To facilitate clear visualization of the cortical vasculature, in-skull field-of-view masking was applied. The PACT images obtained using this virtual imaging pipeline are presented through the case study in Section~\ref{sec:study}.

\section{Case Study: Impact of Skull Modeling Errors}
\label{sec:study}
To demonstrate the utility of the proposed framework in transcranial PACT studies, a case study was conducted to investigate the impact of skull modeling errors. For this case study, the initial pressure distribution simulated with the seven-beam illumination system in Section \ref{sec:exvit} was employed. Unlike the examples in Fig. \ref{fig:prop}, the acoustic-elastic NHPs constructed using Model 2a were used to provide a potentially realistic representation of the diplo\"e. The high resolution of the adjunct CT data in this study enabled the use of this model. Acoustic wave propagation and data acquisition were then simulated as described in Section \ref{sec:exvit}, yielding noisy PACT measurement data.

With the simulated noisy pressure data, image reconstruction was performed under three different acoustic-elastic modeling assumptions: Model 2 with \textbf{true} properties, Model 3, and Model 4. The reconstruction using the true acoustic and elastic properties with the optimization-based reconstruction method (OBRM)~\cite{poudel_iterative_2020} served as the optimal case. For reconstructions assuming Model 3 and Model 4, a gradient-free joint reconstruction method~\cite{huang_gradient-free_2025} was employed. This method jointly estimates the $p_0$ distribution and wave speeds to mitigate inaccuracies arising from mismatches in the assumed properties during reconstruction. It is formulated as a bilevel optimization problem. In the inner optimization, the $p_0$ distribution is estimated given the current wave speed estimates using a penalized least squares objective with total variation regularization. In the outer optimization, the wave speeds are updated by minimizing the mean-squared error between the reconstructed and true $p_0$ distributions within the cortical region. By leveraging the knowledge of the true $p_0$ distribution, this approach provides an upper bound on the best achievable image quality for a given skull model, which is not feasible in experimental studies.

The results of the case study are illustrated in Fig. \ref{fig:casestudy}. The first column displays the true $p_0$ distribution for the entire NHP (top) and the cortical region (bottom). When using the true wave speed maps, the reconstructed $p_0$ distribution closely matches the true $p_0$ distribution in both the scalp and cortical regions, indicating effective aberration compensation. In contrast, assuming Model 3, which treats the diplo\"e as a homogeneous layer with properties distinct from other skull plates, introduced artifacts and noticeable image degradation due to modeling mismatch. This degradation is even more pronounced in the case of Model 4, which assumes the entire skull as homogeneous. 

A critical observation is that the artifacts resulting from these models are highly correlated with the shape of the scalp vasculature, indicating that the error arising from the imprecise skull modeling in the scalp region can propagate into and degrade image quality in the cortical region. Given that the $p_0$ intensity of scalp vasculature is often orders of magnitude stronger than that of cortical vasculature, such artifacts can lead to more severe degradation than the cortical signals themselves. Therefore, assessing aberration compensation without realistic scalp structures and practical optical illumination can lead to misleading evaluations of reconstruction method's performance. The use of realistic NHPs generated by the framework enables systematic and highly valid assessments of such methods.

\section{Discussion}
\label{sec:discussion}
This study presents a 3D NHP generation framework designed to advance virtual imaging studies for transcranial PACT, and is particularly well-suited for supporting data-driven image reconstruction approaches.
The framework is capable of generating NHP ensembles with variations in both the scalp and cortical vasculature, optical properties of blood vessels, and the skull's viscoelastic properties. When ensembles of CT data are available, additional variations in the skull geometry and composition can also be introduced. The framework's capability to generate stochastic ensembles of realistic NHPs enables rigorous assessments of reconstruction algorithms.

The framework's flexibility accommodates diverse research needs by offering skull models with varying levels of heterogeneity. This adaptability allows researchers to select a model that balances their requirements for accuracy, computational resources, and the resolution of available adjunct data.
Among the four models, Model 1, derived entirely from adjunct CT images, provides the most visually realistic skull representation. However, the limited exploration and validation of the assumed linear relationship between HU and elastic and acoustic properties~\cite{angla_transcranial_2023} restricts the model's overall accuracy and applicability. Multiple relationships between CT and viscoelastic properties have been proposed~\cite{aubry_experimental_2003, connor_unified_2002, rho_relations_1995, marsac_ex_2017, vyas_predicting_2016, clement_non-invasive_2002, mcdannold_elementwise_2019} but have not been fully validated. Inaccurate fine-scale modeling of such heterogeneity may yield less reliable predictions of skull-induced aberrations than coarse approximations. Importantly, the framework allows for future customization of these models through the incorporation of more validated relationships or the use of alternative modalities, such as UTE MRI~\cite{deininger-czermak_evaluation_2022}.

The scalp and its vasculature were stochastically synthesized in this work. However, when PACT pressure data from a targeted patient's head are available, the anatomy of tissues external to the skull can be accurately determined through image reconstruction rather than synthesis. This is possible because the pressure data from these regions are less affected by skull-induced aberrations, as demonstrated in the case study presented in this work. Utilizing anatomy derived from the reconstructed PACT images produces phantoms that more closely mimic the target subject, facilitating the fine-tuning of the related image reconstruction methods for subject-specific applications.

A case study on imaging artifacts induced by skull modeling errors was conducted to demonstrate the application of the proposed NHP generation framework. Beyond assessing skull model heterogeneity in image reconstruction, the generated stochastic NHPs can be utilized to create high-quality training datasets for learning-based methods. For instance, the stochasticity in acoustic and elastic property values facilitates the development of a learning-based image reconstruction method~\cite{huang_fast_2025} that addresses inter-subject variability in properties. Furthermore, the generated NHPs may enable statistically rigorous approaches to account for uncertainty in the skull viscoelastic properties values and their spatial heterogeneity by using the Bayesian approximation error approach \cite{kaipio2007statistical,kaipio2013bayesian}. The phantoms are also valuable for investigating the impact of other physical factors, such as skull registration errors.

\section{Conclusions}
\label{sec:conclusion}
This study presents a novel framework for generating ensembles of realistic 3D NHPs based on adjunct CT data to support virtual imaging studies of transcranial PACT. By incorporating skull models with customizable heterogeneity, stochastically synthesized cortical and scalp vasculature, and tissue-specific stochastic assignment of optical and viscoelastic properties, the framework produces phantoms that reflect anatomical and physiological variability. 
The utility of the proposed framework was demonstrated through a case study that investigated the impact of skull modeling errors on PACT image quality. The findings underscored the importance of accurate skull models in image reconstruction. The presented framework establishes a foundation for virtual imaging studies, enabling rigorous evaluation and development of reconstruction methods, including data-driven approaches, and objective assessment and optimization of transcranial PACT imaging system designs. 
To facilitate the immediate application of this work by the research community, a dataset of 50 NHPs has been made publicly accessible. This resource directly supports transcranial PACT studies and advances its translation into clinical uses.

\section*{Acknowledgment}
This research used the Delta advanced computing and data resource which is supported by the National Science Foundation (award OAC 2005572) and the State of Illinois. Delta is a joint effort of the University of Illinois Urbana-Champaign and its National Center for Supercomputing Applications. This work was supported in part by the National Institutes of Health under Awards  R01 EB031585, R01 EB034261 and P41 EB031772.
This work was also supported in part by National Institutes of Health grants U01 EB029823 (BRAIN Initiative), R01 CA282505, and UG3DA065155, as well as grant 2024-337784 from the Chan Zuckerberg Initiative DAF, an advised fund of the Silicon Valley Community Foundation. L.W. has a financial interest in Microphotoacoustics, Inc., CalPACT, LLC, and Union Photoacoustic Technologies, Ltd., which, however, did not support this work.

\bibliographystyle{ieeetr}
\bibliography{references, references-2}

\end{document}